\documentclass[10pt,conference]{IEEEtran}

\usepackage[top=0.8in, bottom=1in, left=0.635in, right=0.635in]{geometry}
\usepackage{cite}
\usepackage{amsmath,amssymb,amsfonts}
\usepackage[flushleft]{threeparttable}
\usepackage{array,booktabs,makecell}
\usepackage{algorithmic}
\usepackage{graphicx}
\usepackage{textcomp}
\usepackage{float}
\usepackage{mathtools}
\usepackage{xcolor}
\usepackage{diagbox}
\usepackage{slashbox}
 \usepackage{graphicx}
 \usepackage[bookmarks=false]{hyperref}

\def\BibTeX{{\rm B\kern-.05em{\sc i\kern
-.025em b}\kern-.08em
    T\kern-.1667em\lower.7ex\hbox{E}\kern-.125emX}}
\begin{document}

\title{
DNN-Based Nulling Control Beam Focusing for Near-Field Multi-User Interference Mitigation 
}

\author{\IEEEauthorblockN{Mohammadhossein Karimi, Yuanzhe Gong, Tho Le-Ngoc}
\IEEEauthorblockA{{Department of Electrical and Computer Engineering, McGill University, Montreal, QC, Canada} \\
Email: mohammad.karimirozveh@mail.mcgill.ca, yuanzhe.gong@mail.mcgill.ca, tho.le-ngoc@mcgill.ca}
}

 \maketitle
 
\begin{abstract}
This paper proposes a deep learning-based framework for near-field nulling control beam focusing (NCBF) in extra-large MIMO (XL-MIMO) systems to mitigate multi-user interference (MUI). A dual-estimator architecture comprising two fully connected deep neural networks (FCDNNs) is developed to separately predict the phase and magnitude components of NCBF weights, using locations of both desired and interfering users. 
The models are trained on a large dataset generated via a Linearly Constrained Minimum Variance (LCMV) beamforming algorithm to accommodate diverse user configurations, including both collinear and non-collinear scenarios.
Illustrative results demonstrate that the proposed DNN models achieve high prediction accuracy, with test errors of only 0.067 radians for phase estimation and 0.206 dB for magnitude estimation.
Full-wave simulations incorporating realistic element radiation patterns and inter-element coupling confirm the close agreement between the beam patterns produced by the DNN-predicted and LCMV-based NCBF schemes under practical deployment conditions.
An average MUI suppression of 36.7 dB is achieved, with interference mitigation exceeding 17.5 dB across all tested cases.
The proposed approach enables scalable and real-time beam focusing with effective interference suppression, offering a promising solution for future near-field multi-user wireless communications.
\end{abstract}


\section{Introduction}
\label{sec:intro}
Extra-large antenna arrays (XLAAs) play a pivotal role in next-generation wireless communication by substantially enhancing both network capacity and reliability \cite{cui2022near,cui2022channel}. 
By incorporating a large number of antenna elements, XLAAs can effectively shape highly directional beamforming that precisely targets users and minimizes interference. 
This spatial multiplexing capability allows multiple users to be served simultaneously, leading to enhanced spectrum efficiency and high data rates. 
Expanding the array aperture leads to a quadratic growth of the radiative near-field region, where a spherical wavefront is established. In contrast to the far-field region, which is limited to angular domain beamforming, XLAA near-field beam focusing precisely controls beam gain in both angular and distance domains, thereby improving spatial isolation in multi-user (MU) communication systems \cite{selvan2017fraunhofer, cui2022channel}. 
This location-differentiated beam focusing enables Location Division Multiple Access (LDMA), potentially boosting spectral efficiency by up to ten times compared to conventional far-field methods \cite{wu2023multiple}.
To leverage the potentials offered by the radiative near-field, a novel distance-parameterized angular-domain sparse model alongside joint dictionary learning and sparse recovery channel estimation methods for controlling XLAA is proposed in \cite{zhang2023near}. 
Similarly, \cite{zhang2022beam} incorporates spherical wavefront phase modelling and accounts for variations in signal amplitude and projected aperture across array elements with various beamforming schemes.
The simulation results demonstrate the effectiveness of the new degree of freedom for interference suppression through distance separation and improvement in sum rate for extra-large multiple-input multiple-output (XL-MIMO) systems.
Meanwhile, most existing XL-array studies \cite{selvan2017fraunhofer, cui2022channel, wu2023multiple, zhang2023near, lu2021near, zhang2022beam} rely on theoretical spherical wavefront models with omnidirectional element pattern assumption while neglecting realistic electromagnetic (EM) interaction between elements. These factors, however, are critical in radiative near-field evaluations, where the coupling waves and heterogeneous element gains can vary rapidly.
To address this gap, \cite{gong2025near} employs iterative full-wave simulations to illustrate the evolution of the radiative near-field with increasing measurement distance. 

The enlargement of array sizes significantly escalates the computational burden associated with optimizing beamfocusing vectors.
To address the challenge, intelligent learning-based techniques can be developed to predict optimal beamformers in real-time by capturing the complex relationships between channel conditions and beam pattern metrics \cite{liu2024near,gong2025near,liu2022deep,monemi20246g,mallioras2022novel,jiang2023near}.
In \cite{liu2024near}, a three-phase graph neural network (GNN)-based beam training scheme is proposed for XL-MIMO systems. By adopting near-field channel models, specialized XL-MIMO beamforming codebooks, and exploiting the position correlation among adjacent users, the approach maps far-field wide beams to near-field codewords, allocates beams based on GNN outputs, and designs a hybrid transmit beamforming scheme.
Similarly, \cite{liu2022deep} presents a deep learning-based beam training method to address the increased pilot overhead introduced by near-field codebooks, utilizing received signals from far-field wide beams. 
In \cite{monemi20246g}, a novel smart spot beamforming strategy is introduced for Fresnel-zone operation. This channel-state information (CSI)-independent DRL approach adaptively concentrates radiated power at a desired focal point, with a scalable design featuring multiple sub-arrays and individual TD3-DRL optimizers to manage computational complexity.

However, similar to other XL-array studies, the evaluation process heavily relies on theoretical near-field propagation models, leaving a noticeable gap in capturing more practical radiation patterns.
Additionally, existing studies \cite{liu2024near, gong2025near, liu2022deep, monemi20246g, mallioras2022novel, jiang2023near} primarily focus on near-field beamforming algorithms that maximize desired signal gain, while overlooking the more challenging objective of minimizing MUI through spatial isolation design using nulling control beam focusing (NCBF) approaches.
Although in \cite{mallioras2022novel}, a recurrent neural network (RNN) architecture is trained to maximize the array gain difference among various users, the approach is restricted to far-field angular differentiation. 
Later, \cite{gong2025near} employs a deep neural network (DNN) to predict near-field NCBF vectors. However, the study is restricted to collinear user scenarios, which represent a considerably less complex user distribution and involve relatively simple prediction challenges.
Consequently, the more general case of near-field user distributions—including both collinear and non-collinear users—remains an open research challenge.


Building on the insights above, a dual-estimator fully connected deep neural network (FCDNN) architecture is proposed, one estimator for phase and the other for magnitude, to predict XLAA NCBF vectors. 
The model is trained on an extensive dataset generated using a Linearly Constrained Minimum Variance (LCMV) NCBF approach, accommodating MU scenarios in which users can be both collinear and non-collinear. 
Tests and comparisons with the reference LCMV beamformer and full-wave simulations confirm that the proposed model accurately and effectively controls both beam focusing and nulling locations.
The remainder of this paper is organized as follows. Section II introduces the near-field phased array system model and details the NCBF algorithms. Section III describes the training data preparation, DNN architecture, and hyperparameter tuning process. Section IV presents the test loss results and full-wave simulation outcomes for various desired and interfering user locations. Finally, Section V concludes the paper and outlines directions for future research.

\section{System Model}
\label{sec:system-model}
\subsection{Near-Field Channel Model}
\label{sec:channel-model}

Based on the distance from the field observation point, the propagation field of the antenna array is typically classified into three main regions: 1) Reactive near-field, 2) Radiative near-field, and 3) Far-field \cite{selvan2017fraunhofer}. 
Given a maximum allowable propagation phase variation of \( \pi/8 \), the  Rayleigh distance, which defines the boundary between the radiative near-field region and the far-field region, can be estimated as $d_R = \frac{2D^2}{\lambda}$,
where \(D\) and \(\lambda\) denote the aperture size of the array and the operating wavelength, respectively \cite{selvan2017fraunhofer}.  
This study focuses on the radiative near-field region. 

Unlike the far-field regime, where impinging waves are assumed to propagate as planar wavefronts, EM waves in the near-field exhibit spherical wavefronts. Consequently, the array steering vector in the near-field depends on both the radial distance and the wave's angle of arrival.
For an \(N\)-element array, the response vector corresponding to a user located at the spherical coordinate \(\boldsymbol{p} = (\phi, \theta, r)\) is expressed as  
\begin{equation}
    \label{eq:channel-model}
\boldsymbol{h}(\boldsymbol{p}) = \left[g_1 e^{j\beta\rho_1},\, g_2 e^{j\beta\rho_2},\, \dots,\, g_N e^{j\beta\rho_N}\right]^T,
\end{equation}
where the term \( e^{ j \beta \rho_m } \) and \(g_n\) denote the phase difference attributable to the varying wave paths travelled by the \( n \)-th element in the array and its corresponding amplitude scaling factor.
\(\beta={2\pi} / {\lambda}\) is the propagation constant. 
The phase shift at each antenna element arises due to variations in the wave propagation path length, which impacts the received signal phase.  
Assuming isotropic element radiation patterns, the channel magnitude gain \(g_n\) can be expressed as  
\begin{equation}
    \label{eq:mag-gain}
    g_n = \frac{\lambda}{4\pi r_n} = \left(\frac{\lambda}{4\pi r_c}\right)\frac{r_c}{r_n},
\end{equation}
where \(r_c\) and \(r_n\) represent the distances between the user at position \(\boldsymbol{p}\) and the center of the antenna array, and the distances between the user and the \(n\)-th antenna element, respectively.  
Furthermore, the free-space path loss between the user and the \(n\)-th antenna element contains a common factor of \(\frac{\lambda}{4\pi r_c}\), which results from the distance between the array center and the user.  
By setting the array center as the phase reference point, a normalized array channel vector can be expressed as  
\begin{equation}\label{eq:channel-model-normalized}
    \boldsymbol{h}'(\boldsymbol{p}) = \left[\frac{r_c}{r_1}e^{j\beta r_1}, \frac{r_c}{r_2}e^{j\beta r_2}, \dots, \frac{r_c}{r_N}e^{j\beta r_N}\right]^T.
\end{equation}

\subsection{Multi-User Interference Suppression System Model}
\label{sec:dnn-system}
In this study, a partially connected hybrid beamforming (HBF) scheme is adopted for the near-field MU XL-MIMO system to enhance hardware efficiency by reducing the number of connections between RF chains and antenna elements \cite{gao2016energy}. 
In this architecture, each RF chain is linked to a subset of antenna elements, forming distinct sub-arrays. This configuration simplifies hardware requirements while preserving effective beamforming capabilities. Moreover, each antenna sub-array generates its own beam pattern, which is controlled by a dedicated beamforming network.


Considering an uplink scenario where a base station (BS) antenna array serves $K$ users in the radiative near-field region, a sub-array with $N$ elements is specifically allocated to the target user located at $\boldsymbol{p_u}$.
The $K-1$ potential interfering users are positioned at $\boldsymbol{p}_k$, where $k=1,...,K$ and $k\neq u$. 
The received signal at the BS is the superposition of signals from the target user and the other $K-1$ interferers. 
After applying the beam focusing weights, the signal can be expressed as
\begin{equation}
    \label{eq:sign-model}
    y = \underbrace{\boldsymbol{w}^T\boldsymbol{h}'(\boldsymbol{p_u})s_u}_{\text{Desired Signal}} + \underbrace{\sum_{k \neq u}\boldsymbol{w}^T\boldsymbol{h}'(\boldsymbol{p}_k)s_{k}}_{\text {Multi-user Interference}}  + \underbrace{\boldsymbol{w}^T \boldsymbol{n}}_{\text {Noise}} ,
\end{equation}
where $\boldsymbol{w} \in \mathbb{R}^{N\times1}$ denotes the beam focusing vector. The terms $s_u$ and $s_{k}$ represent the sending signals from the target user and the $k$-th interfering user, respectively, while $\boldsymbol{n}$ corresponds to additive white Gaussian noise (AWGN). The noise power is given by $\sigma^2$, and the noise is assumed to follow a multivariate Gaussian distribution, expressed as $\boldsymbol{n} \sim \mathcal{N}(\mathbf{0}, \sigma^2 \mathbf{I})$.

Ensuring robust MU system performance necessitates effective suppression of MUI.
Leveraging spatial information, including the angular and distance coordinates of both the desired and interfering users, NCBF plays an important role in mitigating interference power by enforcing radiation nulls at the positions of interfering users, thereby minimizing receiver sensitivity at those locations.

\subsection{Maximum Directivity Beamformer and LCMV NCBF} 
To leverage the advantages of large-scale arrays in shaping radiation patterns, beamforming algorithms play a crucial role by controlling the array’s feeding phase and magnitude. 
Phase alignment filtering effectively adjusts the received signal phases across the array, enabling the design of the maximum directivity beamformer (MDB) at a given location \(\boldsymbol{p}\) by applying the conjugate of the array response vector \(\boldsymbol{h}'(\boldsymbol{p})\).  

Meanwhile, NCBF algorithms can be designed by estimating and inverting the received signal covariance matrix. 
A widely used NCBF approach is the LCMV beamformer \cite{van2002optimum}. 
By enforcing the array response to match a predefined gain vector, the LCMV beamformer preserves the desired signal for a user located at \(\boldsymbol{p}_u\) while forming deep radiation nulls at \(K\) potential interference locations \(\boldsymbol{p}_k\), where \(k \in \{1, \dots, K\}\) and $k\neq u$.  
The LCMV constraint can be formulated as  
\begin{equation}
\label{eq:lcmv_constraints}
\boldsymbol{C}^{\mathrm{H}}\boldsymbol{w}_\mathrm{lcmv} = \boldsymbol{d},
\end{equation}
where \(\boldsymbol{C} = [\boldsymbol{h}'(\boldsymbol{p}_1),\boldsymbol{h}'(\boldsymbol{p}_2), \dots, \boldsymbol{h}'(\boldsymbol{p}_K)] \in \mathbb{C}^{N \times K}\) represents the array response matrix which corresponding to \(K\) users. 
\(\boldsymbol{d} \in \mathbb{R}^{K \times 1}\) denotes the desired gain vector, where the \(u\)th element is set to 1 (i.e., \(\boldsymbol{d}_u = 1\)), while all other elements are 0 (i.e., \(\boldsymbol{d}_k = 0\) for all \(k \in \{1, \dots, K\}\) with \(k \neq u\)).
Given the signal covariance matrix \(\mathbf{R}\), the LCMV beamformer for a $N-$element array is obtained in closed form as \cite{van2002optimum,gong2024perturbation}
\begin{equation}
\label{eq:lcmv_solution}
\boldsymbol{w}_\mathrm{lcmv}(\boldsymbol{p}_1, \dots, \boldsymbol{p}_{K}) =
\mathbf{R}^{-1}\boldsymbol{C}\left(\boldsymbol{C}^{\mathrm{H}}\mathbf{R}^{-1}\boldsymbol{C}\right)^{-1}\boldsymbol{d}=\boldsymbol{a} \odot e^{j\boldsymbol{\phi}},
\end{equation}
where $\odot$ denotes element-wise multiplication between two vectors of the same dimensions. \(\boldsymbol{a} = [a_1, a_2, \dots, a_N]^T\) denotes the magnitude vector components, and \(\boldsymbol{\phi} = [\phi_1, \phi_2, \dots, \phi_N]^T\) represents the phase component vector. 

\section{Proposed DNN Approach}
\label{sec:dnn-approach}
To effectively suppress potential interference while maintaining the desired user's signal gain, a DNN-based dual-estimator is trained for NCBF prediction, which manages the non-constant modulus beam focusing weights. This architecture consists of two separate models that predict the phase and magnitude of the beam focusing weights, respectively, based on the input locations of both target and interfering users.

\subsection{Dataset Generation and Preprocessing}
\label{sec:dataset-preprocessing}
For supervised learning model training, we generate the LCMV-based NCBF vectors \(\boldsymbol{w}_\mathrm{lcmv}(\boldsymbol{p}_1, \boldsymbol{p}_2, \dots, \boldsymbol{p}_{K})\) under various combinations of intended and interfering user locations. These vectors serve as the ground-truth labels, while the corresponding user locations are provided as the input features.

\subsubsection{Dataset Generation}
\label{sec:dataset-generation}
Since the array beam resolution and effective aperture size are mainly constrained by the number of antenna elements in the given direction, our study focuses on the uniform linear array (ULA) with effective beam focusing in the 2D polar coordinates \((\theta, r)\) \footnote{The proposed framework can be readily extended to three-dimensional scenarios when a two-dimensional ELAA is available and covering effective beam focusing \((\phi, \theta,r)\) as input features.} as DNN inputs. 
In this study, a \(1 \times 24\) ULA operating at 3.5 GHz is considered with an inter-element spacing of 4 cm ($\approx 0.47 \lambda$), resulting in a Rayleigh distance of $d_R=19.8$ m. 
The setup is considered to have a consistent array of configurations as our ELAA prototype in the lab and also the full-wave simulation setup. 
As noted in \cite{kosasih2024finite,gong2025near}, effective near-field distance-dependent beam focusing is typically feasible within approximately $0.1 d_R$, due to the rapid increase in beam depth as the focal distance grows.
To ensure that the generated data samples are free from improper user placements and effectively capture rapid variations in the radiative near-field, the user distance from the array center is restricted to the range 0.5 m to 6 m (approximately \(0.025 d_R\) to \(0.3d_R\)). 
Due to the limited radiative near-field region, a three-user scenario consisting of one intended user and two interfering users is considered.
To ensure unbiased learning and improve model generalization, user locations are randomly sampled, preventing the models from overfitting to specific spatial distributions.

\subsubsection{Data Preprocessing}
\label{sec:preprocessing}
To improve training convergence and model accuracy, we normalize the input features by scaling the angular components and user distances to \(\frac{\pi}{2}\) radians and 6 m, respectively, which corresponds to the largest angular location and distance in the input data, ensuring consistent feature ranges. These normalized values are then combined to form the input feature vectors.

Given that NCBF weights are inherently complex, their magnitude and phase components are separately extracted from the LCMV algorithm to facilitate prediction using DNN estimators. 
To eliminate potential phase ambiguity caused by the periodicity of the phase, all phase values are mapped to the range $[0, 2\pi)$ via modulo-$2\pi$ normalization. 
Subsequently, each phase value is adjusted by subtracting \(\phi_1\), the phase of the NCBF weight for the first antenna element, to standardize the phase reference. This step directs the model to learn relative phase variations among antenna elements rather than their absolute values.
To maintain unit-power NCBF weights, the magnitude components derived from the LCMV algorithm are normalized by a scaling factor of $\frac{1}{\sqrt{\sum_{n=1}^N a_n^2}}$, where $a_n$ is the magnitude of the NCBF weight for $n$-th element in the $N$-element array. Moreover, the normalized magnitudes are converted into dB scale to improve numerical stability and to accelerate the convergence of the training objective function.

\subsection{Loss Function for Training and Evaluation}
\label{sec:training-criteria}
The primary objective in developing the DNN models is to minimize the discrepancy between the predicted phase and magnitude values and the LCMV-based NCBF approaches.
For the predicted magnitude vector \(\boldsymbol{\hat{a}} = [\hat{a}_1, \hat{a}_2, \dots, \hat{a}_N]^T\) and the corresponding LCMV ground truth \(\boldsymbol{a}\), the root mean square error (RMSE) is employed to quantify the deviation between the predicted and ground truth magnitude values, which can be expressed as
\begin{equation}
    \label{eq:RMSE}
    J_{a}(\boldsymbol{a}, \boldsymbol{\hat{a}}) = \sqrt{\frac{1}{N} \sum_{n=1}^{N} (a_n - \hat{a}_n)^2}.
\end{equation}

Phase values are inherently periodic and defined modulo-\(2\pi\). As a result, a standard absolute error metric can be misleading when evaluating differences between predicted and true phases, particularly in cases where the difference spans the boundary between \(0\) and \(2\pi\).
To address this, the circular mean absolute error (CMAE) is employed to evaluate the predicted phase vector \(\boldsymbol{\hat{\phi}} = [\hat{\phi}_1, \hat{\phi}_2, \dots, \hat{\phi}_N]^T\) against its LCMV-generated counterpart \(\boldsymbol{\phi}\) as 
\begin{equation}
    \label{eq:CMAE}
    J_{\phi}(\boldsymbol{\phi}, \boldsymbol{\hat{\phi}}) = \frac{1}{N} \sum_{n = 1}^{N} \min\left( \left| \hat{\phi}_n - \phi_n \right|, \, \left|2\pi - \left| \hat{\phi}_n - \phi_n \right| \right| \right),
\end{equation}
where the shortest phase distance is used for error calculation. Adopting these metrics in both training and evaluation ensures that the predicted values accurately capture the magnitude and phase of the NCBF weights.

\subsection{Model Structure}
\label{sec:model-structure}
\begin{figure}[!t]
\vspace{-0.3cm}
    \centering
    \includegraphics[width=0.9\linewidth]{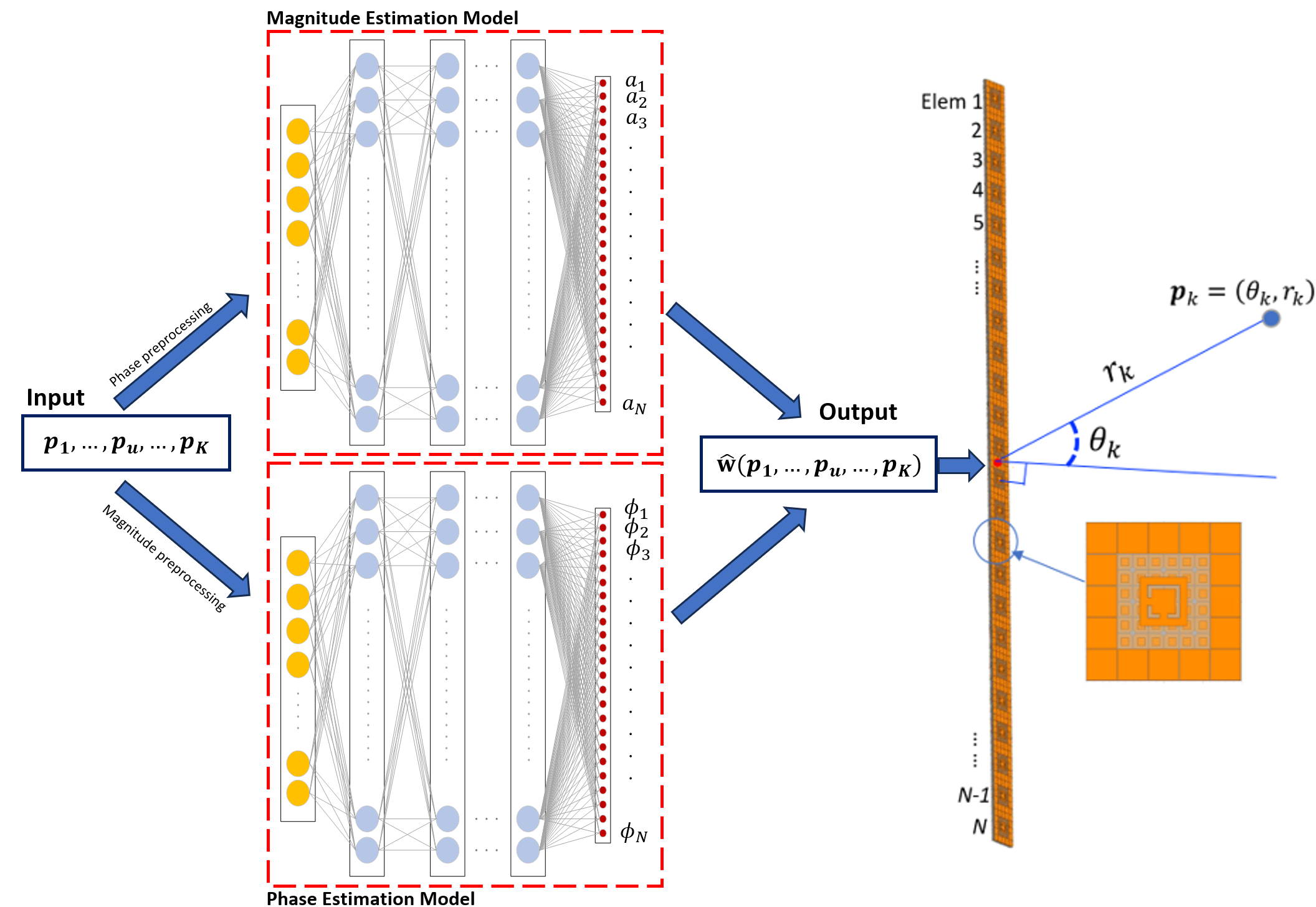}
    \caption{Proposed DNN models for the prediction of NCBF vectors.}
    \label{fig:ِDNN_structure}
    \vspace{-0.2cm}
\end{figure}

As illustrated in Fig.~\ref{fig:ِDNN_structure}, the proposed predictor model consists of two FCDNNs, each utilizing Rectified Linear Unit (ReLU) activation functions for the hidden layers to introduce non-linearity and a linear activation function for the output layer. 
The input layer receives the stacked coordinates of both the intended and interfering users, while the output layers predict the phase or magnitude components of the NCBF vectors. 
The \( l \)-th hidden layer contains \( n_{A,l} \) neurons for the magnitude estimators and \( n_{P,l} \) neurons for the phase estimators.
To predict the NCBF vector for an \(N\)-element array serving \(K\) users concurrently, each DNN follows an architecture of the form \([2K, n_1, \dots, n_L, N]\). 
Here, the input layer contains \(2K\) neurons representing the coordinates ($\theta,r$) of the desired user and the \(K-1\) interfering users, while the output layer comprises \(N\) neurons, each corresponding to one element of the NCBF vector.

\subsection{Hyperparameter Tuning}
\label{sec:hp-tuning}
Optimizing the hyperparameters of DNN models is essential for achieving optimal training performance.
In this experiment, the network architecture, characterized by the number of hidden layers and neurons per layer, is the key parameter governing model complexity, capacity, and computational demand.

In the hyperparameter tuning process, the models were trained on a smaller dataset of 100,000 samples in the three-user scenario, with the data split into 80\% for training and 20\% for validation. 
To evaluate the performance of the DNN models, we experimented with 33 distinct network architectures, each varying in the number of hidden layers (ranging from 4 to 6) and varying in the number of neurons in each hidden layer.
The training process was performed using a batch size of 1024, a learning rate of 0.01, and a total of 150 epochs.  
Each experiment was repeated five times, and the models were ranked based on the average validation loss.
The top-four-performing architectures for the magnitude estimator and phase estimator, along with their corresponding validation loss values for phase and magnitude prediction, are summarized in TABLE~\ref{table:Mag_HP} and~\ref{table:Phase_HP}, respectively.
The best-performing DNN models, A4 for the magnitude estimator and P4 for the phase estimator, were selected, achieving a validation loss of 0.533 dB for magnitude and 0.142 radians for phase, respectively.

\begin{table}[!t]
\vspace{-0.2cm}
\caption{Validation Loss (RMSE) for DNN-based magnitude estimator} \label{table:Mag_HP}
\centering
\normalsize
    \resizebox{\linewidth}{!}{ \begin{tabular}{c|c|c}
   \hline
    Index &Hyperparameter Tuning Model Structure & \makecell{Validation RMSE \\{[dB]}}\\
    \hline
    A1&\makecell{[6, 512, 1024, 1024, 1024, 1024, 512, 24]} &  \makecell{0.569} \\
    \hline
    A2&\makecell{[6, 512, 1024, 1024, 1024, 512, 24]} & \makecell{0.565}\\
    \hline
    A3&\makecell{[6, 1024, 1024, 1024, 1024, 1024, 24]} & \makecell{0.559}\\
    \hline
    A4&\makecell{\textbf{[6, 1024, 1024, 1024, 1024, 1024, 1024, 24]}} & \makecell{\textbf{0.533}}\\
    \hline
    \end{tabular}}
    \vspace{-0.3cm}
  \end{table} 

  \begin{table}[!t]
  \vspace{-0.2cm}
  \caption{Validation Loss (CMAE) for DNN-based phase estimator} \label{table:Phase_HP}
  \normalsize
  \centering
    \resizebox{\linewidth}{!}{ \begin{tabular}{c|c|c} \hline
    Index & Hyperparameter Tuning Model Structure & \makecell{Validation CMAE \\ {[radians]}} \\
    \hline
    P1&\makecell{[6, 1024, 1024, 1024, 1024, 24]} &  \makecell{0.146} \\
    \hline
    P2&\makecell{[6, 512, 1024, 1024, 1024, 512, 24]} & \makecell{0.146}\\
    \hline
    P3&\makecell{[6, 512, 1024, 1024, 1024, 1024, 512, 24]} & \makecell{0.145}\\
    \hline
    P4&\makecell{\textbf{[6, 1024, 1024, 1024, 1024, 1024, 1024, 24]}} & \makecell{\textbf{0.142}}\\
    \hline
    \end{tabular}}
    \vspace{-0.3cm}
  \end{table} 
  
\section{Illustrative Results}
The best-performing architectures achieved in the hyperparameter tuning process for magnitude and phase estimation are employed to train models that predict the NCBF weights' magnitude and phase components.
The final models are trained on an extensive dataset of 1,000,000 user configurations with three-user scenarios, while
80\% of the dataset is allocated for training and 20\% is used for testing.
The phase and magnitude prediction models are trained using a batch size of 1024 and initial learning rates of 0.01 and 0.001, respectively. 
To accelerate convergence toward optimal objective function values, an exponential decay learning rate scheduler with a decay factor of 0.99 per epoch is applied.

\subsection{DNN Model Training and Test Loss}
\label{sec:training}
\begin{figure}[!t]
\vspace{-0.3cm}
    \centering
\includegraphics[width=0.9\linewidth]{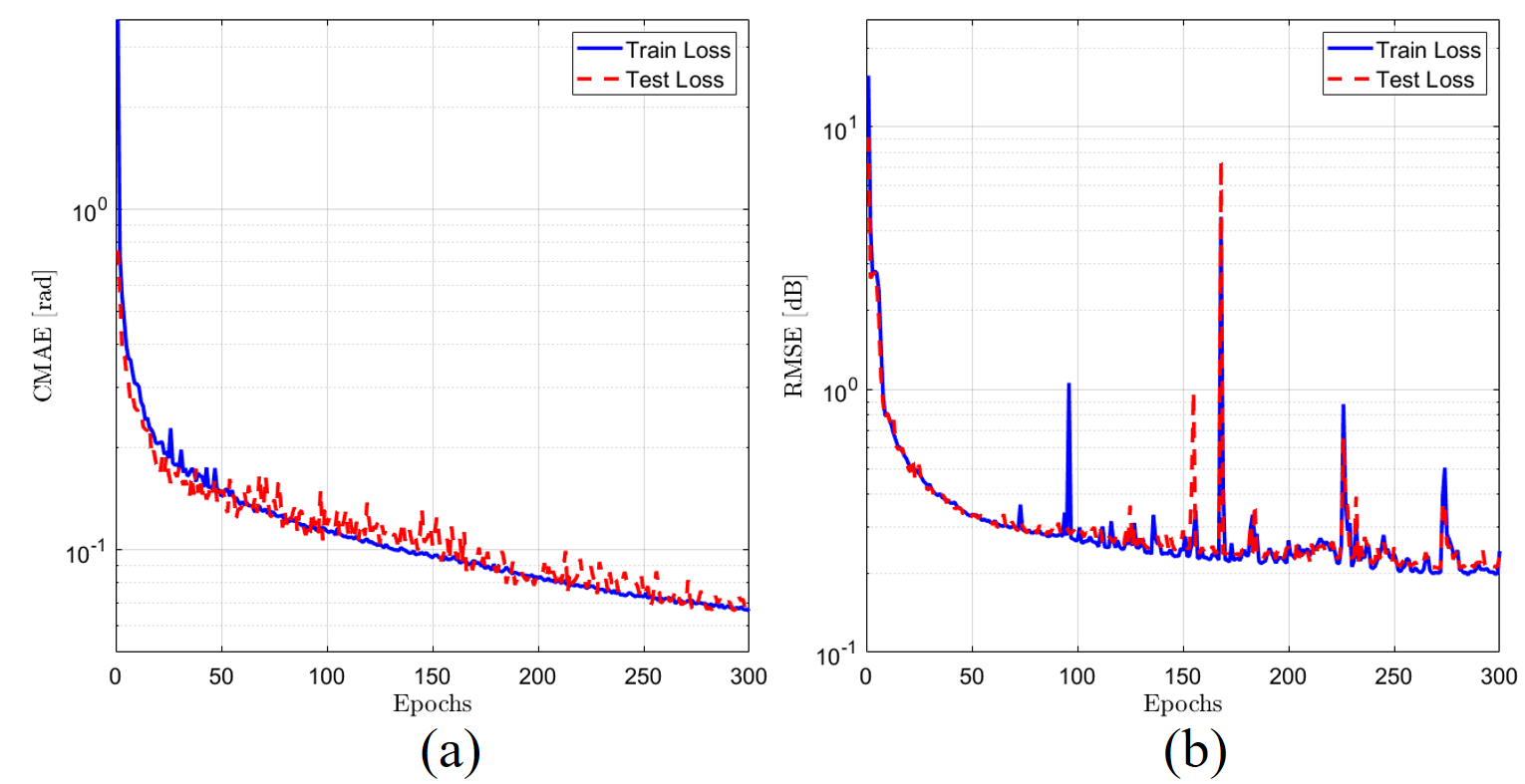}
    \caption{DNN training and test losses versus epochs for (a) Phase prediction model in radians and (b) Magnitude prediction model in dB.}
    \label{fig:ِDNN_model}
    \vspace{-0.1cm}
\end{figure}
The performance of the optimized DNN models for predicting NCBF vectors is evaluated by tracking the training and test losses over 300 epochs. The loss trajectories for both magnitude and phase predictions are illustrated in Fig.~\ref{fig:ِDNN_model}. 
The best-achieved training and test losses for the phase and magnitude estimators are summarized in Table~\ref{table:train-test-loss}.
Fig.~\ref{fig:ِDNN_model}~(a) shows the convergence of the CMAE for phase prediction, with both training and test losses decreasing steadily and following similar trends throughout the epochs. 
The losses converge to values below 0.1 radians, with the final test loss reaching 0.067 radians, indicating high accuracy in phase estimation.
Similarly, in Fig.~\ref{fig:ِDNN_model}~(b), the RMSE in dB for magnitude prediction also shows consistent convergence behavior. 
While the loss curves exhibit consistent downward trends, occasional spikes are likely due to the higher sensitivity of loss function to larger gradients towards the end of training. 
Both training and test losses eventually stabilize at a low RMSE of 0.21 dB, demonstrating the model’s robust performance in predicting magnitude components.
The convergence behavior and low final loss values indicate strong generalization capability and effective learning. The proposed DNN-based dual-estimator reliably predicts both phase and magnitude of the NCBF weights, enabling accurate near-field beamforming control.

\subsection{Full-Wave Simulation with DNN Predicted NCBF Vector}
\begin{figure}[!t]
\centerline{\includegraphics[width=0.85\linewidth]{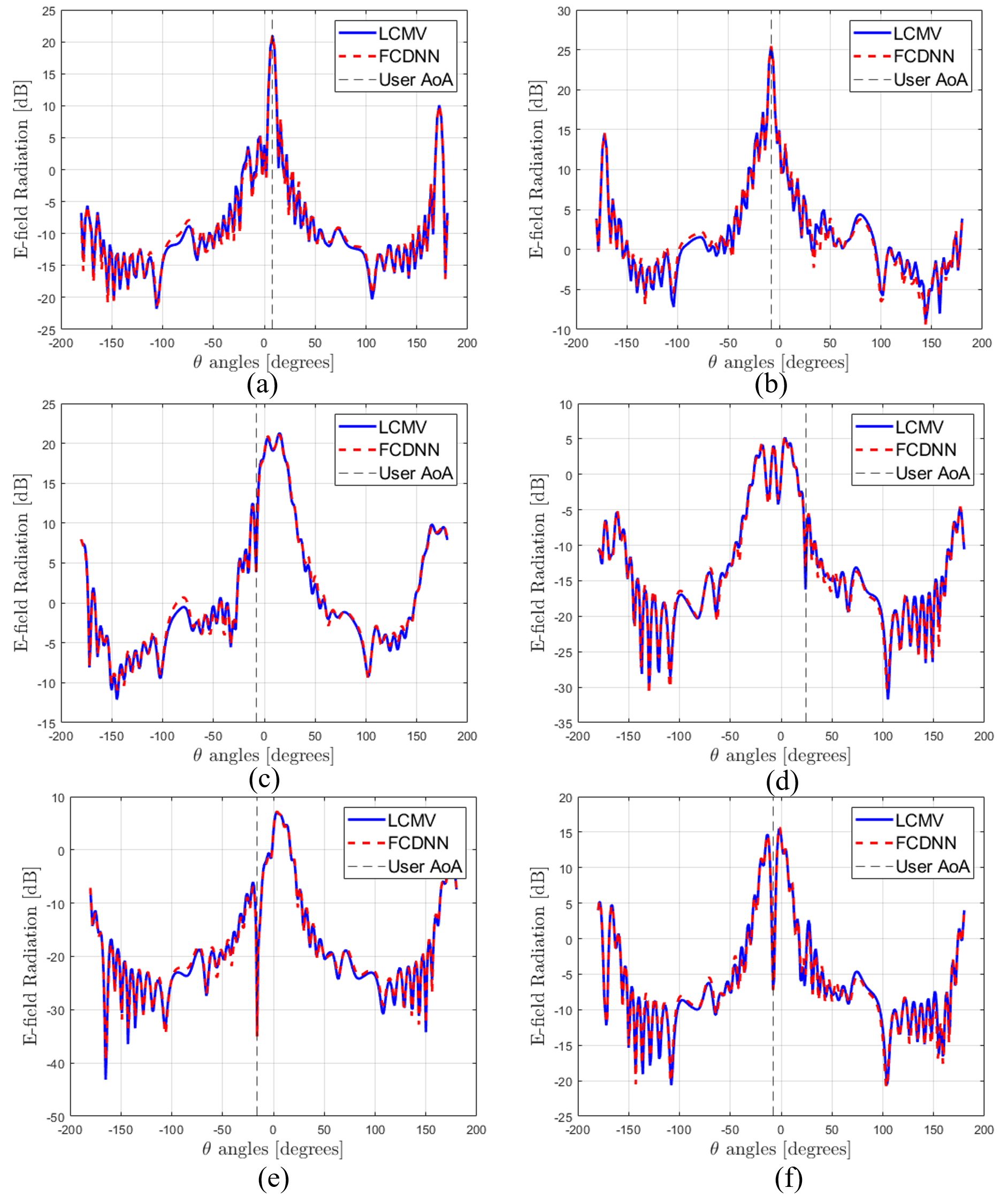}}
\vspace{-0.4cm}
\caption{Ansys HFSS $1\times24$ patch array simulated radiation pattern 2D cuts at the corresponding user distance for two sample three-user cases. Scenario 1: desired user at (a) $\boldsymbol{p}_1= (8^\circ, 1.6\,\mathrm{m})$, interfering user at (c) $\boldsymbol{p}_2= (-8^\circ, 0.8\,\mathrm{m})$, and (e) $\boldsymbol{p}_3= (-16^\circ, 4.9\,\mathrm{m})$; Scenario 2: desired user at (b) $\boldsymbol{p}_1= (-8^\circ, 0.85\,\mathrm{m})$, interfering user at (d) $\boldsymbol{p}_2= (24^\circ, 4.5\,\mathrm{m})$, and (f) $\boldsymbol{p}_3= (-8^\circ, 1.8\,\mathrm{m})$.}
\label{HFSS_DNN_3UE}
\vspace{-0.4cm}
\end{figure}

 \begin{table}[!t]
 \vspace{-0.3cm}
  \caption{Magnitude and Phase Prediction Models Train and Test Loss} \label{table:train-test-loss}
  \normalsize
  \centering
    \resizebox{0.8\linewidth}{!}{ \begin{tabular}{c|c|c}
     \hline
    Estimator & \makecell{Training Loss} & \makecell{Test Loss}\\
    \hline
    \makecell{Phase Estimator} &  \makecell{0.066 radians} & \makecell{0.067 radians} \\
    \hline
    \makecell{Magnitude Estimator} & \makecell{0.202 dB} & \makecell{0.206 dB}\\
    \hline
    \end{tabular}}
     \vspace{-0.3cm}
  \end{table} 
To assess the effectiveness of the trained DNN models, a \(1 \times 24\) metamaterial-based patch antenna array was designed and simulated in Ansys HFSS, with an inter-element spacing of 4 cm, as illustrated in Fig.~\ref{HFSS_DNN_3UE}~(g). Comprehensive details of the patch element design are presented in \cite{gong2022miniaturized}.
Floquet port setup was used to enable precise control over the feeding weights of each element.
The NCBF vectors predicted by the trained DNN model were applied to two representative three-user near-field scenarios operating at 3.5 GHz. 
The resulting near-field radiation patterns were analyzed. The first scenario, shown in Fig.~\ref{HFSS_DNN_3UE}~(a) to (c), involves users spatially separated in both angle and distance within the near-field region. The second scenario, depicted in Fig.~\ref{HFSS_DNN_3UE}~(d) to (f), demonstrates the system’s capability to handle the configuration comprising both collinear and non-collinear users.

The resulting beam patterns demonstrate that the NCBF weights predicted by the DNN-based estimator closely align with those generated by the LCMV algorithm, achieving accurate control of both main-lobe steering and interference nulling. 
With the DNN-predicted NCBF vector, a minimum and average beam focusing gain difference of 17.5 dB and 36.2 dB is achieved between the intended and interfering users in the tested three-user scenarios, enabling effective MUI suppression. The achievable beam focusing gain difference with the DNN-predicted vector shows a less than 3 dB difference compared to that obtained using the LCMV approach.

To comprehensively evaluate the effectiveness of the DNN-generated NCBF weights, MATLAB-based simulations using the Phased Array Toolbox and a spherical wavefront setup were conducted over 10,000 randomly generated samples. 
Two key performance metrics were assessed: (i) NCBF gain, defined as the ratio of beam focusing gain at the target user’s location to that at the interfering user’s location, and (ii) angular deviation, which measures the offset between the actual interference source and the corresponding null direction.
The results show that the proposed DNN model achieves an average NCBF gain of 36.7 dB between the target and interfering users. Furthermore, the average angular deviation in null placement is less than $0.5^\circ$, demonstrating the model’s high accuracy and effectiveness in suppressing interference.

\subsection{Time Complexity Analysis}
The forward pass of a FCDNN with $L$ layers generally has an asymptotic time complexity of $\mathcal{O}\!\left(\sum_{l=1}^{L-1} n_l n_{l+1}\right)$ per sample. Furthermore, by employing a batch size $B$, which is chosen according to the memory and parallelization limits of the computation device, the time required to process $M$ samples is upper bounded by $\mathcal{O}\!\left(\tfrac{M}{B}\sum_{l=1}^{L-1} n_l n_{l+1}\right)$. Considering that the size of the largest layer in the DNN is $n_{\max}$, the batch prediction complexity can be simplified to $\mathcal{O}\!\left(\tfrac{M}{B}n_{\max}^2\right)$. Thus, the prediction time complexity of NCBF weight generation scales quadratically with the largest layer size of the model.

For the LCMV algorithm, let $N$ and $K$ denote the number of antenna elements and the number of users, respectively. For each sample, computing the covariance matrix requires $\mathcal{O}(N^2)$ operations, inverting the covariance matrix requires $\mathcal{O}(N^3)$, and computing the NCBF weights requires $\mathcal{O}(N^2K + K^3)$. Therefore, the overall time complexity per sample can be summarized as $\mathcal{O}(N^2 + N^3 + N^2K + K^3)$. This indicates that the complexity of the LCMV grows cubically with the number of antennas and the number of users.

\begin{figure}[H]
\centering
\vspace{-0.2cm}
\includegraphics[width=0.65\linewidth]{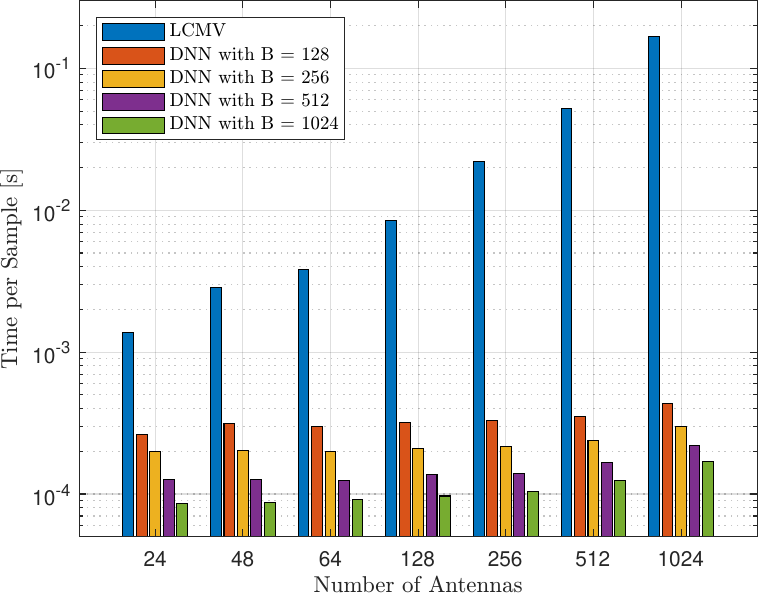}
\caption{Average computation time of NCBF weights: LCMV vs. DNN models.}
\label{fig:time_complexity}
\end{figure}

To evaluate the computational intensity and time efficiency of the proposed method against LCMV, NCBF weights for 10,000 samples are computed using both the dual-DNN model and the LCMV algorithm. The DNN models are tested with different batch sizes, illustrating the achievable computation times under varying hardware resources. 
All experiments are conducted on an Intel Xeon Gold 5220R CPU, and the average time per operation is reported in Fig.~\ref{fig:time_complexity}. 
The results show that the computation-time advantage of DNNs over conventional LCMV grows significantly with increasing array size, while DNN prediction times remain below 0.5 ms in all tested cases. In contrast, the runtime of the LCMV algorithm increases rapidly with the number of antenna elements, leading to excessive delays in ELAA systems with more than 64 antennas.

\section{Conclusions}
This paper presented a DNN-based near-field NCBF framework for mitigating MUI in XL-MIMO systems. To address the challenges of real-time beam pattern optimization in near-field scenarios, a dual-estimator FCDNN was proposed to separately predict the magnitude and phase components of the NCBF weights.
Comprehensive hyperparameter tuning was conducted to identify the optimal model structure, balancing prediction accuracy with model complexity.
The final model achieved a CMAE of 0.067 radians and an RMSE of 0.206 dB in predicting the phase and magnitude of NCBF weights on the test dataset.
Extensive evaluations, including full-wave EM simulations using Ansys HFSS, confirmed that the performance of DNN-predicted NCBF weights closely matches that of the LCMV solution in maintaining gain at the target user location while effectively suppressing interference with minimal angular deviation. The comparison of time complexity demonstrates that the proposed DNN models significantly outperform the LCMV algorithm in terms of time efficiency, with the performance gap widening for ELAA systems.
\vspace{-0.1cm}

\bibliographystyle{IEEEtran}
\bibliography{bibliography.bib}

\end{document}